\documentclass[longbibliography, twocolumn, superscriptaddress,nofootinbib]{revtex4-1}

\usepackage{graphicx}  
\usepackage{dcolumn}   
\usepackage{amssymb, amsmath}
\usepackage[utf8]{inputenc}
\usepackage{hyperref}
\usepackage{float}

\usepackage{amsfonts}

\usepackage{color}

\newcommand{\Lya}{Ly-$\alpha$ }

\newcommand{\units}[1]{{{\rm \ #1}}}

\DeclareUnicodeCharacter{2212}{-}

\begin{document}

\title{\bf{Probing New Physics at Cosmic Dawn with 21-cm Cosmology}}

\author{Omer Zvi Katz}
\affiliation{School of Physics and Astronomy, Tel-Aviv University, Tel-Aviv 69978, Israel}

\author{Nadav Outmezguine}
\affiliation{Berkeley Center for Theoretical Physics, University of California, Berkeley, CA 94720, U.S.A.}
\affiliation{Theory Group, Lawrence Berkeley National Laboratory, Berkeley, CA 94720, U.S.A.}

\author{Diego Redigolo}
\affiliation{INFN, Sezione di Firenze, Via Sansone 1, 50019 Sesto Fiorentino, Italy}

\author{Tomer Volansky}
\affiliation{School of Physics and Astronomy, Tel-Aviv University, Tel-Aviv 69978, Israel}

\begin{abstract}

\noindent
21-cm cosmology provides an exciting opportunity to probe new physics dynamics in the early universe.  In particular, a tiny sub-component of dark matter that interacts strongly with the visible sector may cool the gas in the intergalactic medium and significantly alter the expected absorption signal at Cosmic Dawn. 
However, the information about new physics in this observable is obscured by astrophysical systematic uncertainties. In the absence of a microscopic framework describing the astrophysical sources, these uncertainties can be encoded in a bottom up effective theory for the 21-cm observables in terms of  unconstrained astrophysical fluxes.  In this paper, we take a first step towards a careful assessment of the degeneracies between new physics effects and the uncertainties in 
these fluxes. We show that the latter can be constrained by combining measurements of the UV luminosity function, the Planck measurement of the CMB optical depth to reionization, and an upper bound on the unresolved X-ray flux.  Leveraging those constraints, we demonstrate how new physics signatures can be disentangled from astrophysical effects. Focusing on the case of millicharged dark matter, we find sharp predictions, with small uncertainties within the viable parameter space.\footnote{This paper is a contribution to the proceeding of the Nobel Symposium on Dark Matter.}  
\end{abstract}
\maketitle

\section{Introduction}

Cosmological observables such as the Cosmic Microwave Background (CMB) and the galaxy power spectrum provide invaluable evidence for the existence of dark matter (DM), pointing 
to a pressureless fluid that feebly interacts with the Standard Model. Furthermore,  analysis of the same cosmological data gives detailed information regarding possible non-trivial dynamics of a so-called dark sector in which DM could reside~\cite{Ivanov:2020ril,DePorzio:2020wcz,Xu:2021rwg,Cyr-Racine:2015ihg,Archidiacono:2022iuu,Bottaro:2023wkd,Dvorkin:2013cea,Gluscevic:2017ywp,DAmico:2020kxu,Xu:2018efh,Rogers:2023ezo}.   
 Within this wealth of information from the early universe, 21-cm cosmology is tracking the behavior of the neutral hydrogen in the inter-galactic medium (IGM) at different redshifts~\cite{Pritchard:2011xb}. 
At $z\lesssim6$, 21-cm intensity mapping provides a new tracer of cosmic structure~\cite{Peterson:2006bx} which can complement optical surveys of Large Scale Structure~\cite{CosmicVisions21cm:2018rfq}. For  $z\gtrsim6$ these measurements provide a unique opportunity to probe the (rather dark) Universe since recombination 
and until the Epoch of Reionization 
-- a period that remains poorly understood and challenging to investigate through alternative observables. Nevertheless, such measurements do not unfold without their inherent difficulties (see for example Ref.~\cite{Pritchard:2011xb,10.1088/2514-3433/ab4a73,Liu:2022iyy} for reviews on the subject).

A measurement far from the EoR, at $z\gtrsim 30$, is accompanied by severe experimental challenges which may be addressed by the futuristic program of lunar interferometry~\cite{Liu:2022iyy}. Conversely, in order to understand measurements at $6\lesssim z\lesssim 30$, during the so-called Cosmic Dawn, non-linear structure formation and stellar evolution must be understood and hence predictions suffer from significant systematic uncertainties, even assuming standard cosmology~\cite{Pritchard:2011xb,10.1088/2514-3433/ab4a73}. As a consequence, utilizing 21-cm measurements to detect or constrain any non-standard dark sector dynamics requires a detailed estimation of these uncertainties. This is the goal of this letter.

The interest in the Cosmic Dawn epoch has been reinvigorated by the EDGES observational result in 2018 which suggested a strong absorption signal in the  global 21-cm spectrum~\cite{Bowman:2018yin}. The  signal was significantly  stronger than the maximal absorption signal possible within standard cosmology, hinting towards non-standard dynamics. 
The EDGES result has since been  disputed by the SARAS3 collaboration~\cite{Singh:2021mxo} and many subtleties in the estimation of its significance might be hiding in the assessment of the systematic uncertainties which are plaguing the global 21-cm spectrum measurements~\cite{Hills:2018vyr,Bradley:2018eev,Sims:2019kro,Pund:2023ceu}. More information is expected to come from high-redshift 21-cm interferometry that should be able to shed light on the global signal results~\cite{HERA:2022wmy}. 

Irrespective of the unsettled experimental status, it is natural to ask: {\it 'What kind of non-standard dynamics could leave detectable imprints in 21-cm observables at Cosmic Dawn?'} In particular, exploring the potential for detecting imprints of a dark sector is intriguing. To date, broadly speaking, five dark sector effects have been identified: 
\begin{itemize}
\item {\bf Dark cooling} models in which the dark sector acts to  cool the gas in the IGM through DM elastic collisions~\cite{Tashiro:2014tsa,Munoz:2015bca,Munoz:2018pzp,Barkana:2018lgd,Barkana:2018qrx,Berlin:2018sjs,Liu:2019knx,Barkana:2022hko}. 
\item {\bf Dark heating},  where the dark sector is heating up the gas through DM annihilations or decays~\cite{Evoli_2014,Lopez_Honorez_2016,Liu_2018, D_Amico_2018, Cheung_2019, Mitridate_2018, Clark_2018, Acharya_2023,facchinetti202421cm}.
\item {\bf Modification of the Raileygh-Jeans tail} via a resonant conversion of dark sector particles to CMB photons~\cite{Pospelov:2018kdh,Bondarenko:2020moh,Caputo:2020avy,Caputo:2022keo}. 
\item {\bf Dark sector effects on the power spectrum.} Such changes may delay or enhance radiation which heats up or ionizes the gas (see e.g.~\cite{Driskell:2022pax}) and therefore affects the 21-cm signal.  
\item {\bf Dark sector effects on star  formation.} In particular, suppressed stellar production~\cite{qin2023birth} or variation in their evolution may impact the 21-cm signal~\cite{Ellis:2021ztw}. 
\end{itemize}

Each of the above categories requires detailed analyses to ensure that  phenomenologically viable models exist.  
More importantly, new physics effects may be degenerate with the inherent astrophysical uncertainties, a prevailing challenge during the Cosmic Dawn era. Within the above categories, certain scenarios exhibit complete degeneracy. 
For example, fuzzy or warm DM models lead to a suppressed matter power spectrum at small scales,  resulting in a reduced density of Pop-III stars. As this stellar population remains observationally elusive, it is currently impossible to distinctly separate the primary effects of such dark sectors using 21-cm cosmology.
The situation, however, is often more optimistic and in this letter we aim to take an initial stride in quantifying the degeneracies, concentrating specifically on the dark cooling category outlined above.


The theoretical challenges of realizing a  viable dark cooling model have been addressed in the specific example of Ref.~\cite{Liu:2019knx,Barkana:2022hko},  where a tiny fraction of the DM energy density (below the current CMB bound~\cite{dePutter:2018xte,Boddy:2018wzy}) is  millicharged. If the millicharged component interacts with the rest of the cold DM through a long range interaction, the heat capacity of the cooling bath is enhanced at low redshift, thereby enhancing the signal at Cosmic Dawn. One therefore concludes that 21-cm observables have the opportunity to probe a genuinely unexplored region of the millicharge DM parameter space which lies below the current accelerator constraints~\cite{Liu:2019knx,Davidson:2000hf,Badertscher:2006fm,Prinz:1998ua,Magill:2018tbb,chatrchyan2013search,ArgoNeuT:2019ckq}, but couples too strongly to be probed in regular terrestrial direct detection experiments~\cite{Liu:2019knx,Emken:2019tni}.

Given this well defined theoretical setup, we focus on the challenge of identifying the underlying degeneracies  between the dark cooling signal and astrophysical uncertainties. 
In particular, the  dark cooling acts to enhance the absorption signal at Cosmic Dawn well beyond the one predicted within standard cosmology.  We would therefore like to ask in which region of the parameter space a measurable enhancement  could occur \emph{independently} of the astrophysical uncertainties. The detailed exploration of our analysis, including improved constraints on the astrophysical parameters,  is presented in Ref.~\cite{our} where the different possible stellar formation models and the independent code developed to evaluate their implications on the global 21-cm spectrum are described. 


Our paper is organized as follows. In Sec.~\ref{sec:21cmbasic} we give a brief introduction to the 21-cm global signal, discussing both standard cosmology as well as dark cooling.     In Sec.~\ref{sec:astro} we identify the important astrophysical functions, which constitute an effective theory for 21-cm cosmology, and review the relations between them and the simplified astrophysical models.   In the process we   determine the corresponding astrophysical uncertainties and discuss existing and upcoming constraints.    In Sec.~\ref{sec:results} we summarize some of our results.    Further details and results are presented in~\cite{our}.


\section{21-cm with Dark Cooling}~\label{sec:21cmbasic}
In this section we introduce the basics of 21-cm cosmology in Sec.~\ref{sec:basics}, focusing on the sky-averaged 21-cm brightness and its behavior at Cosmic Dawn. Then, we illustrate the evolution of the IGM temperature in the presence of dark cooling in Sec.~\ref{sec:TK&DC}.
\subsection{21-cm Basics}\label{sec:basics}
The CMB photons measured today have propagated the universe since redshift of order $1100$, traversing cold neutral hydrogen clouds before reaching us. During this time, photons with a wavelnegth of $21$-cm were repeatedly  absorbed and emitted through the hyperfine transitions of ground state hydrogen. The resulting deviation in the tail of the CMB spectrum as measured today is usually parametrized by the differential 21-cm brightness temperature~\cite{1958PIRE...46..240F,1959ApJ...129..551F,1997ApJ...475..429M,Pritchard:2006sq,Mesinger:2010ne}
\begin{align}\label{eq:T21}
    T_{21}(z) &= \frac{\left(T_s(z)-T_\gamma(z)\right)}{1+z}\left(1-e^{-\tau_{21}(z)}\right) \\
    & \simeq 27x_{\text{HI}}\left[1-\frac{T_\gamma}{T_s}\right]\!\left[\frac{1+z}{10}\frac{0.15}{\Omega_m h^2}\right]^{1/2}\!\left[\frac{\Omega_b h^2}{0.023}\right]\,\text{mK}\,,\notag
\end{align}
where $T_\gamma$ is the CMB temperature and $T_s$ is the effective spin temperature, quantifying the relative occupation of the two hyperfine levels of the hydrogen ground state. The second equality is obtained by assuming that the optical depth of 21-cm photons remains small, i.e. $\tau_{21}\ll1$. In the absence of new physics, this condition is satisfied until the end of reionization, where the neutral fraction of hydrogen, $x_{HI}(z)\equiv n_{HI}(z)/n_H(z)$\footnote{Throughout this work we use the standard notation to distinguish between the different baryonic components. In this notation: HI, HeI stand for neutral hydrogen and helium atoms, HII and HeII represent singly ionized hydrogen and helium atoms, and finally HeIII are doubly ionized helium atoms. When writing H or He it is to be understood as as the total hydrogen and helium populations, including all levels of ionization.} drops significantly.  

The dynamics of $T_s$ is set by 
spin flipping processes: i) Resonant absorption and emission of 21-cm CMB photons. This process is controlled by the induced emission and absorption rates $B_{10}=B_{01}/3$. ii) Hydrogen scatterings with helium, the residual ionized fraction and other hydrogen atoms. We express the collisional de-excitation rate as $C_{10}=\sum_{i=\text{HI,e,p}} n_ik^i_{10}$, where $n_i$ are the number densities of the different species and the $k^i_{10}$ factors depend on the kinetic temperature of hydrogen, $T_{\rm k}$, and are tabulated in Ref.~\cite{Furlanetto:2006su,Zygelman_2005,Allison1969,Pritchard:2006sq}. iii) Absorption of \Lya photons followed by a de-excitation to the opposite spin state (also known as the Wouthuysen Field (WF) effect~\cite{1958PIRE...46..240F,1952AJ.....57R..31W}). Since \Lya photons are emitted by stars, the corresponding de-excitation rate, parameterized by $P_{10}$, will depend on detailed astrophysics. 
    
Throughout the evolution of $T_s$, the dominant spin-flipping rate is always much grater than the Hubble expansion rate, driving the system into a steady state given by
\begin{equation}\label{eq:Ts}
T_s^{-1}=\frac{ T_{\gamma}^{-1}+\bar x_\alpha T_{\alpha}^{-1} +\bar x_k T_{k}^{-1}}{1+\bar x_\alpha+\bar x_k}\,,
\end{equation}
where the collisional and \Lya coupling coefficients are defined as $\bar{x}_k = C_{10}/B_{10}$ and $\bar{x}_\alpha = P_{10}/B_{10}$, and the \Lya temperature, $T_\alpha$, is an effective color temperature defined through the equilibrium relation of the WF spin flip interactions, i.e. $P_{01}/P_{10} = 3e^{-\frac{E_{21}}{T_\alpha}}$. 

For completeness we also write the WF spin flipping rate in its common parametrization~\cite{Hirata:2005mz}
\begin{equation}
P_{10} = 4\pi \int \sigma_{10}(E)J_\alpha(E) dE\equiv  6\pi\lambda_\alpha^2\gamma_\alpha \tilde{S}_\alpha  \bar{J}_\alpha \,, 
\label{eq:P10}
\end{equation}
where $\sigma_{10} = \frac{3}{2}\lambda^2_\alpha\gamma_\alpha\phi_{10}(E) $ is the WF de-excitation cross section, $\lambda_\alpha$ is the wavelength of \Lya photons, $\gamma_\alpha=0.74\alpha_{\text{em}} E_\alpha^3a_0^2\simeq4.3\times 10^{-7}\text{ eV}$ is the line half-width at half-maximum, and $\phi(E)$ is the corresponding line profile. $J_\alpha$ is the specific number intensity (number of photons per time, energy, area and steradian) around the \Lya energy, resulting from the emission of UV photons by early luminous sources. 

The spectrum around the \Lya line center is modified by energy transfer between \Lya photons and hydrogen atoms as a result of recoil and spin exchange in WF interactions \cite{Chen:2003gc,Hirata:2005mz}. This modification, together with the details of the line profile are encapsulated in $\tilde{S}_\alpha = \int \phi_{10}(E) J_\alpha/\bar{J}_\alpha$, where $\bar{J}_\alpha$ is defined as the intensity at the red edge of the line and hence unaffected by the WF scatterings. 
In our 21-cm code, we calculate  $\tilde{S}_\alpha$ and $T_\alpha$ following Ref.~\cite{Hirata:2005mz}\footnote{ Ref.~\cite{Hirata:2005mz} assumed $T_\alpha \gg E_{21}$ in deriving $\tilde{S}_\alpha$. Under this assumption $P_{01}=3P_{10}$, and their definition is equivalent to ours.}, while the intensity $\bar{J}_\alpha$ depends on astrophysics and will be discussed in Sec.~\ref{sub:lya}.

At the end of the Dark Ages ($z\sim 30$), $T_s$ is coupled to $T_\gamma$, resulting in $T_{21} = 0$. However, at the onset of the Cosmic Dawn, star formation results in \Lya fluxes which quickly become the dominant spin flipping rate, coupling $T_{\rm s}$ to $T_\alpha$. Prior to heating from astrophysical sources, one has $T_{\rm k}<T_{\alpha}<T_\gamma$ and thus the 21-cm signal is found in absorption. This absorption signal of Cosmic Dawn is the focus of this study. Crucially, the amplitude and width of this absorption signal in standard cosmology depends on the astrophysical fluxes controlling $T_\alpha$, $T_{\rm k}$ and $x_{HI}$ in Eq.~\eqref{eq:T21}. The goal of this study is to disentangle this dependence from new dark sector dynamics affecting $T_{\rm k}$, whose time evolution we describe below.

\begin{figure}[t!]
    \centering
    \includegraphics[width=0.5\textwidth]{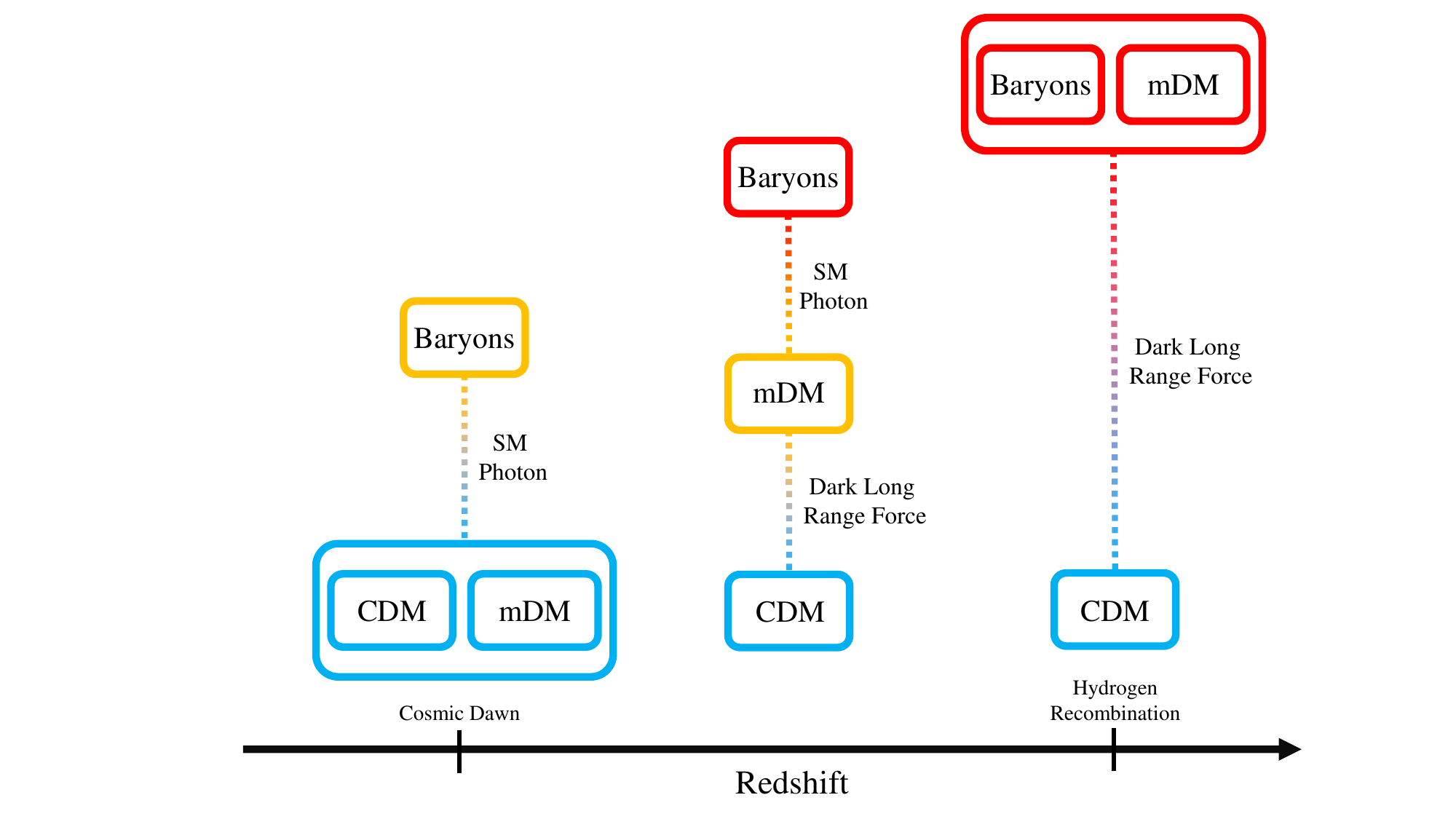}
    \caption{The 2cDM dark sector~\cite{Liu:2019knx} is comprised of two DM components. The majority of DM, assumed to be cold DM (CDM), does not engage in tree-level interactions with SM particles, while the remaining small fraction, $f_\text{m}$, is assumed to posses an electric charge. In addition, the two DM components interact with one another through a new long range dark force. The evolution of the 2cDM model is as follows: i) prior to hydrogen recombination, mDM is strongly coupled to baryons, evading CMB constraints as long as $f_m\le4\times10^{-3}$~\cite{dePutter:2018xte,Boddy:2018wzy}. During this period the long range interactions between CDM and mDM are suppressed due to the large relative velocity. ii) As the latter dissipates, mDM begins to cool through its interactions with CDM, eventually decoupling from the baryons. Acting as a large heat bath, CDM absorbs any excess heat from the mDM, always keeping it colder than the baryons which continuously deposit heat to the mDM fluid through elastic scatterings, mediated by SM photons. iii) In some regions of the 2cDM parameter space, mDM eventually couples to CDM.}
    \label{fig:2cDM_sketch}
\end{figure}

\begin{figure*}[t]
    \centering
    \includegraphics[width=0.8\textwidth]{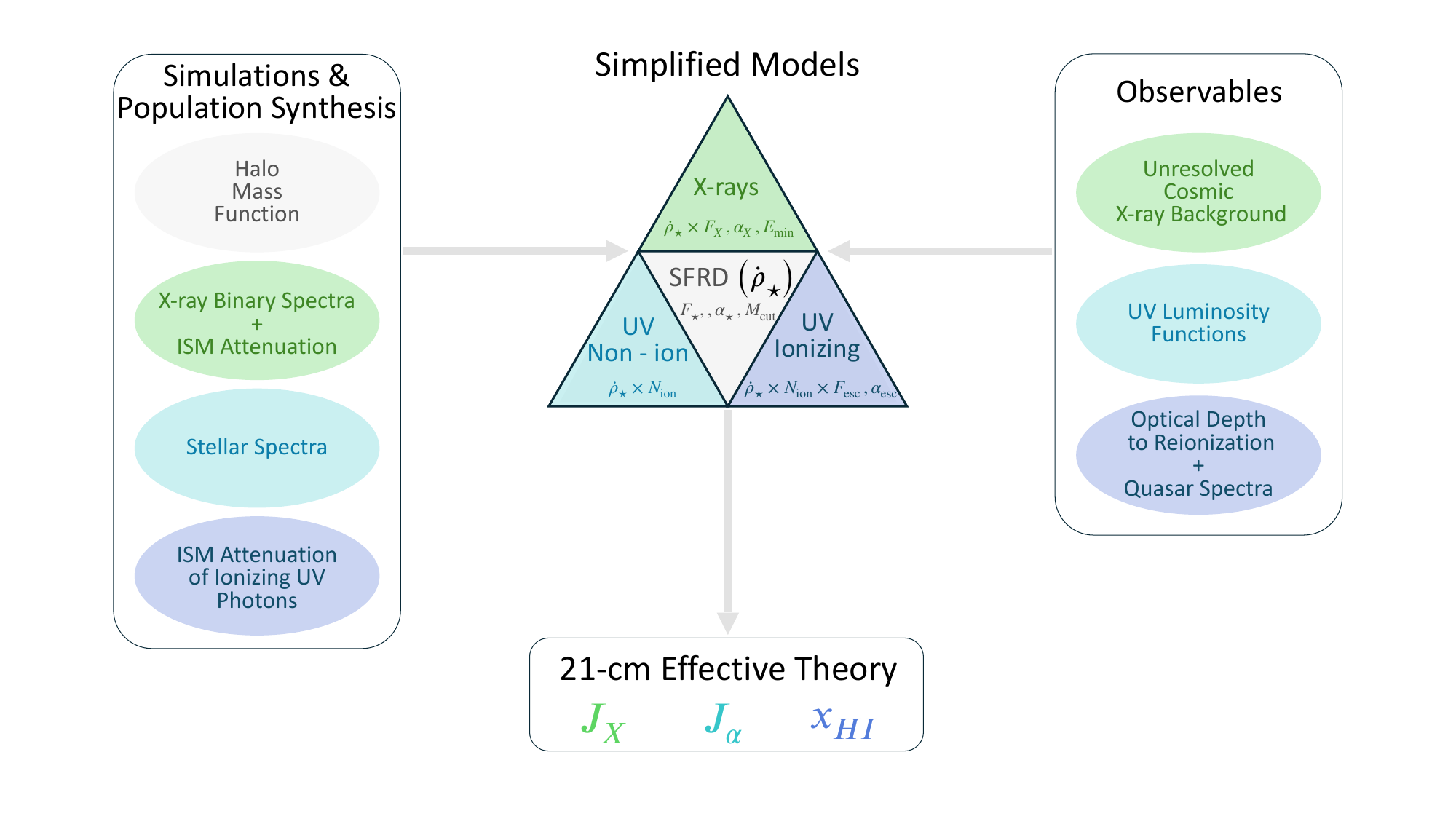}
    \caption{An illustration depicting the constraining procedure of the 21-cm effective theory described by three astrophysical functions:  i) $J_\alpha$ (\textbf{light blue}) - the specific intensity at the \Lya line, ii) $J_{\rm X}$ (\textbf{green}) - the specific intensity at X-ray energies, and iii) $x_{\rm HI}$ (\textbf{purple}) - the neutral fraction of hydrogen. These functions may be characterized using simplified astrophysical models represented as triangles with corresponding colors. All three models rely on yet another simplified model of star formation (central \textbf{light grey} triangle).  The parameters of each simplified model are listed beneath its title.  On the left side we list the data extracted from simulations and population synthesis models, which are integrated into the construction of the simplified models. On the right side, the measured observables used to constrain each simplified model are outlined, maintaining the same color coding.}
    \label{fig:AstroSketch}
\end{figure*}

\subsection{The Kinetic Temperature of Hydrogen and Dark Cooling}
\label{sec:TK&DC}
Including both SM and BSM contributions, the kinetic temperature of baryons evolves as
\begin{eqnarray} 
    \frac{dT_{k}}{d\log a} = -2T_K 
    &+&\frac{2}{3H}\Big(\dot{Q}_{\text{Compton}} + \dot{Q}_X 
    \nonumber \\ 
    &&+ 
    \dot{Q}_{\text{Ly}\alpha} + \dot{Q}_{\text{CMB}} + \dot{Q}_{\text{BSM}} \Big) \, ,
\label{eq:Tk evolution}
\end{eqnarray}
where $a$ is the scale factor, $H(z)=\sqrt{\Omega_m} H_0(1 + z)^{3/2}$ is the Hubble rate controlling the adiabatic cooling, and the $\dot{Q}_i$s represent the heat transfer rates per baryon as a result of different SM and BSM processes which we briefly discuss.

At high redshifts, hydrogen is mostly ionized, CMB photons continuously Compton scatter with the residual electron fraction, coupling $T_\text{k}$ to $T_\gamma$ through $\dot{Q}_{\rm Compton}$. However, eventually hydrogen recombines and adiabatic cooling prevails. The era of adiabatic cooling (Dark Ages) only terminates once first stars form, emitting X-ray and UV radiation.  The former propagates to the IGM, heating it through photoionization,
while the latter redshifts or cascades to the \Lya energy where the photons exchange heat with the hydrogen atoms as a result of WF scatterings. Additionally, scatterings of $21$-cm CMB photons and hydrogen atoms result in spin flipping interactions affecting the hydrogen kinetic temperature which is encoded in $\dot{Q}_{\text{CMB}}$.

The heat transfer rates $\dot{Q}_\text{X}$, and $\dot{Q}_{\text{Ly}\alpha}$ strongly depend on the details of the astrophysical processes, including the formation of stellar objects and their emission properties. This dependence is encapsulated in specific number intensities for X-rays, $J_\text{X}$, and for \Lya photons, $\bar{J}_\alpha$. The modeling these intensities is described in Sec.~\ref{sec:astro}, but generically we have $\dot{Q}_\text{X}\gg\dot{Q}_{\text{Ly}\alpha}$. 

The expressions for the Compton and X-ray heating rates are common in the 21-cm literature and can be found in Refs.~\cite{Pritchard:2006sq, 2011MNRAS.411..955M}. On the other hand, being significantly smaller, the \Lya and CMB terms are often neglected. However, these rates are enhanced at low $T_\text{k}$, and may therefore play a role in situations with extreme dark cooling and suppressed X-ray flux. We therefore incorporate these rates in our 21-cm code. For calculating $\dot{Q}_{\rm Ly\alpha}$ we follow Ref.~\cite{Chen:2003gc}, but accounting for the full \Lya line profile derived in Ref.~\cite{Hirata:2005mz}. For $\dot{Q}_{\rm CMB}$ we follow Ref.~\cite{Meiksin:2021cuh}, which corrects the original result of Ref.~\cite{Venumadhav:2018uwn}.

Finally, $\dot{Q}_{\text{BSM}}$ in Eq.~\eqref{eq:Tk evolution} accounts for the dark sector contribution to the $T_K$ evolution. Here we focus on the effect of dark cooling through DM-SM elastic scatterings. Specifically, we take as a benchmark dark cooling model the two coupled DM (2cDM) sector, first proposed in Ref.~\cite{Liu:2019knx} whose mechanism is reviewed in Fig.~\ref{fig:2cDM_sketch}. In this work we will assume a benchmark 2cDM model where the amount of dark cooling is maximized: the millihcrage fraction is set at the CMB bound $f_\text{m}=4\times10^{-3}$~\cite{dePutter:2018xte,Boddy:2018wzy}, the cold DM mass is fixed at the BBN bound $m_\text{C}=10\,\rm{
 MeV}$~\cite{Boehm:2013jpa} and the mDM-CDM cross section is fixed to the maximal value allowed by CMB constraints as discussed in Ref.~\cite{Liu:2019knx}. This leaves only two free parameters in the dark sector, the mDM mass, $m_\text{m}$, and charge, $Q$, whose viable parameter space is shown in Fig.~\ref{fig:ConstraintsPrediction}. We leave a more comprehensive study of the 2cDM parameter space to Ref.~\cite{our}.

\section{Astrophysical Uncertainties}\label{sec:astro}

\begin{figure}[t]
    \centering
    \includegraphics[width=0.5\textwidth]{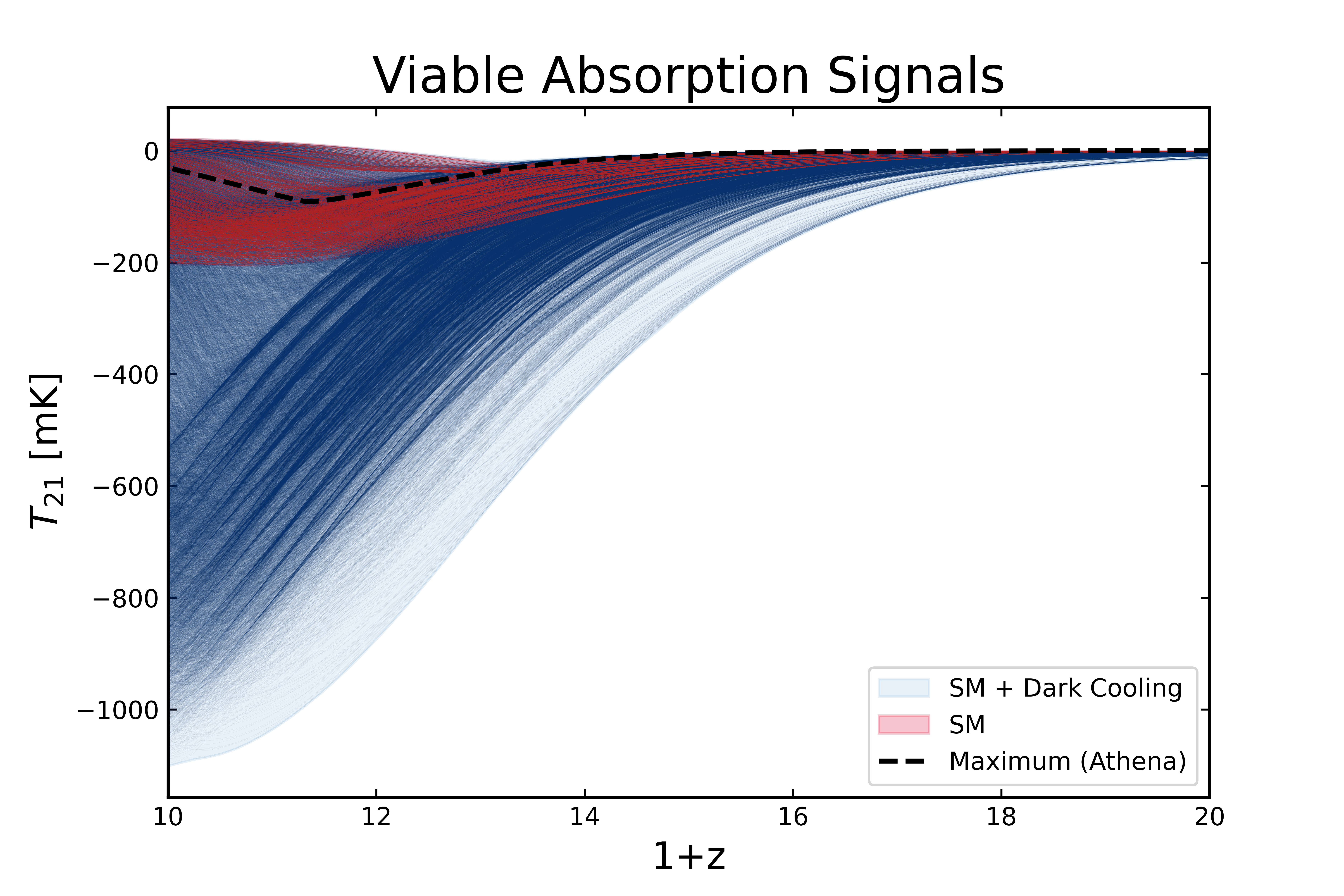}
    \caption{The evolution of the 21-cm global Cosmic Dawn signal assuming only SM physics (\textbf{red}), and including dark cooling (\textbf{blue}), as predicted over the viable astrophysics and 2cDM parameter space. Each line corresponds to a specific combination of astrophysics and dark sector parameters.  Here, we assume the star formation model by Ref.~\cite{Park:2018ljd} as discussed in Sec.~\ref{sec:starform} and constrain the astrophysical fluxes by imposing current measurements of the UV luminosity function~\cite{Park:2018ljd,Bouwens_2015} (Sec.~\ref{sub:lya}), the optical depth to reionization measured by Planck~\cite{Planck:2018vyg} and the measured spectra of bright quasars~\cite{McGreer:2014qwa} (Sec.~\ref{sec:ion}), and the expected constraint on the unresolved X-ray flux from Chandra~\cite{Lehmer:2012ak,Fialkov:2016zyq},  discussed in Sec.~\eqref{sec:Xray}.  The {\bf dashed black} line shows the expected boundary of these models, once results from the future Athena experiment become available~\cite{Marchesi:2020smf}. For concreteness, we fix the mDM-CDM cross section to its maximal value allowed by CMB constraints,  take  the millicharged component relative density to be $f_\text{m}=4\times10^{-3}$, and fix the mass of the CDM component at $m_\text{C}=10$ MeV.}  
    \label{fig:T21_of_z_scan}
\end{figure}

The behavior of the brightness temperature in Eq.~\eqref{eq:T21} depends on three astrophysical functions: i) the \Lya flux $\bar{J}_\alpha$, ii) the X-ray flux $J_X$, and iii) the ionized fraction $x_{HI}$ which depends on the UV and X-ray fluxes together. 
If left completely undetermined, a very strong X-ray flux, a very inefficient \Lya emission, or an early reionization can in principle make the Cosmic Dawn absorption signal arbitrarily flat for any scenario of dark cooling, making any statement about the dark sector dynamics plagued by astrophysical uncertainties.  

 In what follows we show that  using a combination of present and future observables and simulations one 
 can sufficiently bracket the three astrophysical functions to make robust statements about the dark sector with only few assumptions about the spectrum of astrophysical fluxes. The  full procedure we follow to achieve this goal is illustrated in Fig.~\ref{fig:AstroSketch} and  is detailed in the remainder of this section. 

 The final result of this analysis is summarized in Fig.~\ref{fig:T21_of_z_scan} where the envelope of all possible 21-cm brightness temperature evolutions is shown for standard cosmology (red) and with the 2cDM dark cooling scenario (blue). Crucially, current constraints on the UV luminosity function~\cite{Park:2018ljd,Bouwens_2015} and the optical depth to reionization~\cite{Planck:2018vyg,Park:2018ljd} imply that  astrophysical scenarios with very inefficient \Lya emission or early reionization are excluded. We also impose existing and expected constraints on the unresolved X-ray flux from Chandra~\cite{Lehmer:2012ak,Fialkov:2016zyq} and Athena~\cite{Marchesi:2020smf} respectively. The latter is expected to improve the current constraint from Chandra by at least two orders of magnitudes. 

\subsection{Models of star formation}\label{sec:starform}
A basic ingredient in modeling the global intensity of astrophysical sources is the rate of their formation. We focus specifically on the formation of stars, and define the star formation rate density (SFRD) as the rate at which mass is converted to stars per co-moving volume.
The SFRD can be expressed as an integral of the halo mass function (HMF), $dn/dM_h$\footnote{Following the modeling of Ref.~\cite{Park:2018ljd}, instead of the time derivative in Eq.~\ref{eq:SFRD}, we divide the stellar mass by a characteristic timescale $H^{-1}(t)t_\star$, where $t_\star$ is a dimensionless parameter.}:
\begin{equation}\label{eq:SFRD}
    \dot{\rho}_\star^{i}(t)=\frac{d}{dt}\int_0^\infty m^{i}_\star(M_h) \frac{dn}{dM_h}(M_h,t)dM_h\,,
\end{equation}
where the mapping function, $m^{i}_\star(M_h)$, relates the total mass of a given halo, $M_h$, to its stellar mass. Motivated by the  extrapolation of high redshift UV luminosity functions (UVLFs), the star formation efficiency in halos exceeding a threshold virial mass of $M_{\rm cut}$ can be modeled as a powerlaw in $M_\text{h}$~\cite{Bouwens_2015,Park:2018ljd}. Below $M_{\rm cut}$ -- the precise value of which depends on the baryonic composition of the halo and on stellar feedback effects -- star formation is highly suppressed. 

To be specific, we focus on the star formation model described in Ref.~\cite{Park:2018ljd} which considers a single star population (Pop-II stars) described by three parameters: i) the normalization of the star formation efficiency $F_\star$, ii) the power law index $\alpha_\star$, and iii) the cut-off of star formation $M_{\rm cut}$. More realistic star formation models, in particular including an extra population of Pop-III stars~\cite{Munoz:2021psm} will be considered in~\cite{our}.




To evaluate the SFRD in our scenario we use the $\Lambda$CDM HMF with the prescription of ~\cite{2016MNRAS.456.2486D} available in the public python toolkit COLOSSUS~\cite{Diemer_2018}. The cosmological parameters other than $\tau_{\text{reion}}$ are fixed using the $\Lambda$CDM fit to the Planck data~\cite{Planck:2018vyg}, assuming they are unaffected by the dark sector.  
While this assumption is certainly valid for a fraction of mDM with $f_m<4\times 10^{-3}$ it would be interesting to check if in the 2cDM model, discussed in Sec.~\ref{sec:TK&DC}, the late time recoupling of CDM to mDM can lead to a sizeable suppression of small scale structures (see Ref.~\cite{Driskell:2022pax} for a related study).

\subsection{The \Lya intensity}\label{sub:lya}
%


\Lya photons are produced by non-ionizing UV photons either redshifting from below the Ly-$\beta$ line or cascading down through higher Lymann lines. $\bar{J}_\alpha$ of Eq.~\eqref{eq:P10} can then be written in terms of the UV emissivity, $\epsilon_{\rm UV}$, which sums all the astrophysical sources weighted by their probability to end the cascade with the production of a \Lya photon~\cite{Barkana:2004vb,Hirata:2005mz}

During the Cosmic Dawn, stars are the primary source of UV photons~\cite{Wyithe_2005,Barkana:2004vb}, with the most massive stars making the dominant contribution at the Lymann band energies~\cite{Hirata:2005mz}. Assuming that the lifetime of the stars is short with respect to variations in the  star formation rate implies that the volume-averaged comoving number emissivity of UV photons is proportional to the star formation rate~\cite{Barkana:2004vb}
\begin{equation}
     \epsilon_{\rm UV}(E,t) =\frac{\dot{\rho}_\star(t)}{\mu_b}\left\langle\frac{d N_*}{dE}\right\rangle\,, \label{eq:UVEmissivity}
\end{equation}
where $\epsilon_{\rm UV}$ gives the number of UV photons from all sources per unit time, energy and comoving volume, $\mu_b$ is the average baryon mass   and $\left\langle\frac{d N_*}{dE}\right\rangle$ is the average number of photons emitted by a single baryon per unit energy, summed over all relevant stellar populations. 

Since the \Lya flux depends only on the spectra, $\left\langle\frac{d N_*}{dE}\right\rangle$, within a very small energy band, its exact shape  is irrelevant and can be set by its normalization, $N_{\rm ion}$. Consequently, the emissivity in Eq.~\eqref{eq:UVEmissivity} is entirely defined by $N_{\rm ion} \dot{\rho}_\star$. Since $N_\text{ion}$ is degenerate with the normalization $F_\star$ of $\dot{\rho}_\star$, we fix  $N_{\rm ion}=5000$.


Data collected by the Hubble Space Telescope (HST) can be used to map the rest frame UV luminosity function (UVLF) at $4\lesssim z \lesssim 10$~\cite{Bouwens_2015}. These UVLFs can then be used to constrain the parameters associated with stellar formation simplified models ($F_\star$, $\alpha_\star$, $M_{cut}$) once $N_{\rm ion}$ is fixed. The preliminary constraints we use in our analysis were first derived in~\cite{Park:2018ljd,Bouwens_2015}. The James Webb Space Telescope is expected to improve on these constraints in the near future.

\subsection{The X-ray intensity}\label{sec:Xray}
Soft X-ray photons (with energy $E\le2\units{keV}$) continuously interact with the IGM through photoionizations\footnote{X-rays with energy greater than approximately $2\text{ keV}$ have a mean free path longer than the Hubble distance at $z=10$~\cite{Oh:2000zx,McQuinn_2012}. As a consequence, only X-rays with energies below this value play a role in heating and ionizing the IGM.}. Their intensity can be derived by solving the energy transfer equation in a homogeneously absorbing, non-emitting medium~\cite{Mirocha:2014faa}. The modeling of $J_{\rm X}$ can then be related to the global co-moving number emissivity, $\epsilon_{\rm X}$, which depends on the characterization of the X-ray sources. 

Based on observations of nearby starburst galaxies~\cite{Grimm:2002ta,ranalli20032, Gilfanov:2003bd,Fabbiano:2005pj,Mineo:2011id}, it is believed that the dominant X-ray sources during Cosmic Dawn are high-mass X-ray binaries (HMXBs). Being short lived, the emissivity (and luminosity) of the HMXBs population traces the star formation rate, and can therefore be written as in Eq.~\eqref{eq:UVEmissivity}. In accordance with local measurements~\cite{Grimm:2002ta,ranalli20032, Gilfanov:2003bd,Fabbiano:2005pj,Mineo:2011id} and high redshift simulations~\cite{Fragos:2013bfa,Das:2017fys}, the averaged photon spectrum that contributes to the IGM heating can be modelled as a single truncated power law in energy, 
\begin{equation}\label{eq: averaged photon spectrum X}
\left\langle\frac{d N_X}{dE}\right\rangle= \frac{1}{E_0}\left[\frac{E}{E_0}\right]^{\beta_X}\!\!\!\!\Theta \left[E-E_{\mathrm{min}}\right] \Theta \left[E_{\mathrm{max}}-E\right]\, ,
\end{equation}
where  $E_{\mathrm{max}}=2\text{ keV}$ is the maximal energy and the reference energy $E_0$ is typically related to the ratio of the luminosity and star formation rate in the corresponding energy band~\cite{Pritchard:2006sq, Cohen:2016jbh, Park:2018ljd} which we rescale with a normalization parameter $F_\text{X}$.  $E_{\text{min}}$ is the minimal energy that an X-ray photon must posses in order to escape its host galaxy. Its precise value depends on the assumed local density and metallicity of the galaxy~\cite{Das:2017fys}. In our scan we take $E_{\rm min}\in (0.19, 0.85) \text{ keV}$, which 
corresponds to the 2$\sigma$ range in which the  X-rays optical depth in the interstellar medium as extracted from the hydrodynamical simulations of Ref.~\cite{Das:2017fys} is exactly one. Finally, since the variation of $\beta_X$ is less significant than the one of $E_{\rm min}$~\cite{greig2017simultaneously} we fix it to be $\beta_X=-2$, in accordance to high-redshift simulations~\cite{Das:2017fys,Fragos:2013bfa,Park:2018ljd}. 

An upper limit on the X-ray flux can be derived by assuming that early X-ray sources are responsible for the totality of the unresolved cosmic X-ray background (CXB) in the $\left[0.5,2\right]\text{ keV}$ energy band, observed today  by Chandra~\cite{Lehmer:2012ak,Fialkov:2016zyq,Cappelluti:2012rd}. Defining  $z_X^{\text{un}}$ as the maximal redshift above which none of the X-ray sources are resolved by a given telescope, the following constraints from the Chandra measurement can be derived
\begin{equation}
\frac{1}{\left(1+z_X^{\text{un}}\right)^3}\int_{0.5\,\text{keV}}^{2\,\text{keV}}\!\!\! E J_\text{X}\left(E,z_X^{\text{un}}\right) dE < 0.5 \frac{\rm{keV}}{\rm{cm^2\, sec\, sr}}\,. \label{eq:Xraybound}
\end{equation}
Since star formation accelerates at lower redshifts, sources at the lowest redshift dominate the CXB so that a lower $z_X^{\text{un}}$ results in a stronger bound on $F_X$. Given that Chandra reported resolved soft X-ray galaxies only up to $z\sim1$, we conservatively take $z_X^{\text{unres}}=3$.  The upper limit on the right hand side of Eq.~\eqref{eq:Xraybound} will be further improved with the future Athena experiment~\cite{Marchesi:2020smf} which is expected to give a bound at least two order of magnitude tighter.   Since we also expect its resolution to improve, we show the Athena expected limit taking $z_X^{\rm unres} = 4$.

\begin{figure*}[t]
\centering
\includegraphics[width=0.52\linewidth]{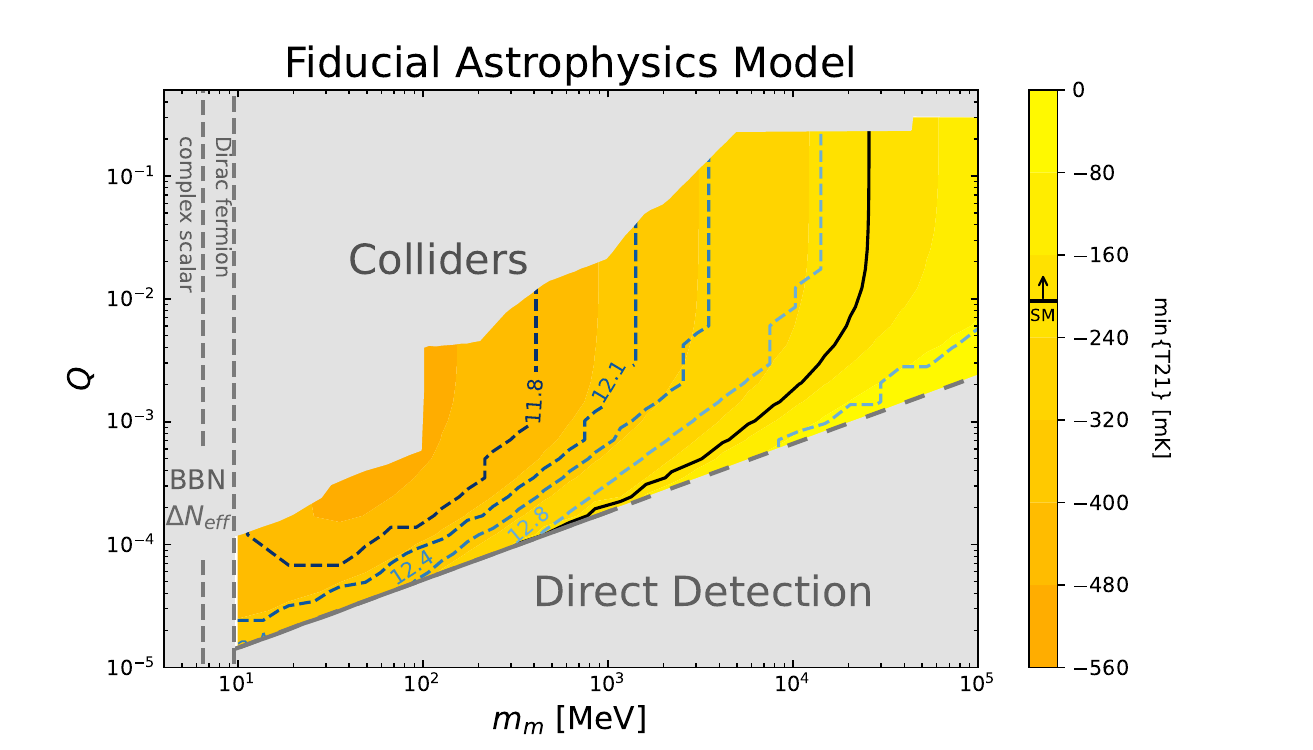}
\includegraphics[width=0.45\linewidth]{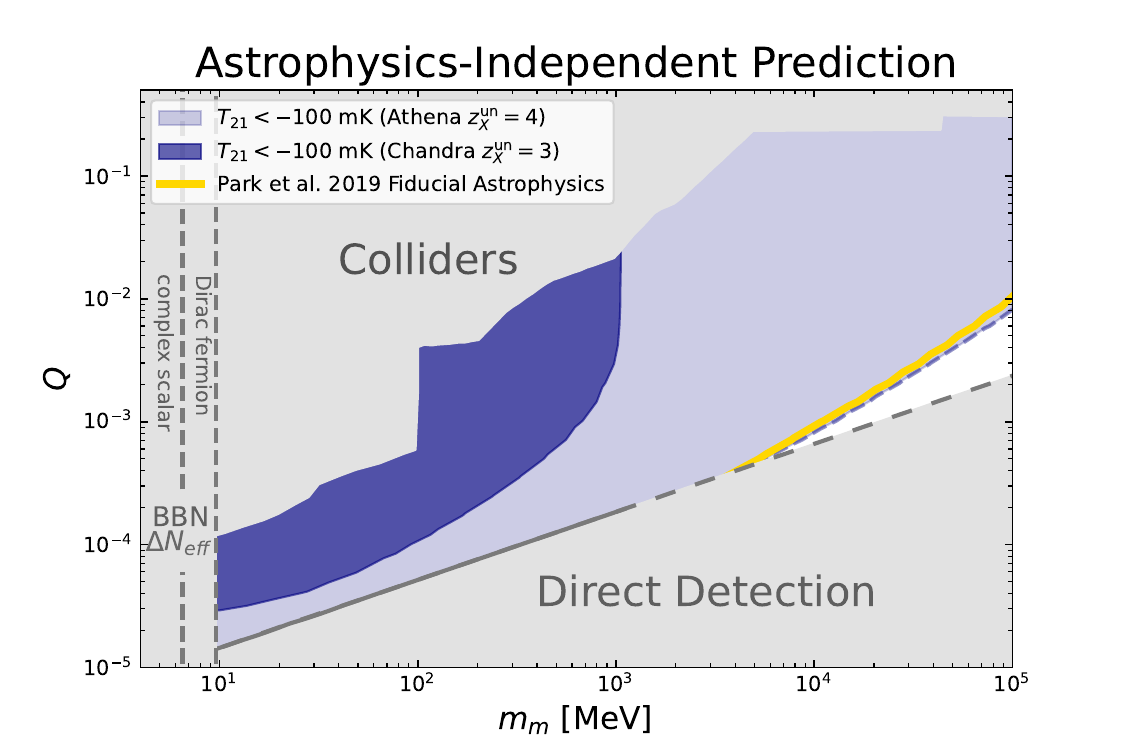}
\caption{The parameter space of millicharged dark matter in the charge $Q$ vs mass $m_m$ plane, for the 2cDM model of Ref.~\cite{Liu:2019knx} (see summary in Fig.~\ref{fig:2cDM_sketch}). The \textbf{grey} shaded regions are excluded by BBN measurements of $\Delta N_{\rm{eff}}$~\cite{Barkana:2018qrx,Munoz:2018pzp,Berlin:2018sjs,Creque-Sarbinowski:2019mcm}, accelerator experiments~\cite{Davidson:2000hf,Badertscher:2006fm,Prinz:1998ua,Magill:2018tbb,chatrchyan2013search,ArgoNeuT:2019ckq}, and direct detection searches collected in~\cite{Liu:2019knx,Emken:2019tni}. CMB constraints on mDM-baryon and mDM-CDM interactions are accounted for in the choice of the 2cDM parameters. \textbf{Left:} The predicted minimal brightness temperature, $T_{21}$, of the Cosmic Dawn absorption signal for the fiducial  astrophysical model of Ref.~\cite{Park:2018ljd} (see Sec.~\ref{sec:results} for the precise choice of parameters). The minimal temperature is indicated in shades of {\bf Yellow} while the {\bf dashed blue} lines specify the redshift at which the minimum is achieved. The {\bf solid black line} on the color bar  marks the minimal $T_{21}$ allowed within the Standard Model for all viable astrophysics models (see also Fig.~\ref{fig:T21_of_z_scan}), while the {\bf solid black contour} on the main figure marks the $Q$, $m_\text{m}$ values needed to reach this temperature assuming the fiducial astrophysics. \textbf{Right:} The astrophysics-independent prediction 
for the expected exclusion assuming a null observation with a −100 mK sensitivity and under the assumptions of a power-law   SFR model fitted to  UVLF data~\cite{Bouwens_2015,Park:2018ljd}. In all of the shown millicharged DM parameter space, a brightness temperature as low as $-100$ mK can be obtained  with some choice of viable astrophysics model. Within the {\bf dark (light) purple} region, $T_{21}<-100\units{mK}$ is predicted  for {\it any} viable astrophysics model that satisfies the constraints discussed in Sec.~\ref{sec:astro}, including the X-ray constraint of Chandra~\cite{Lehmer:2012ak,Fialkov:2016zyq} (Athena~\cite{Marchesi:2020smf}). The {\bf yellow} line shows the predicted  $T_{21}=-100\units{mK}$ for the fiducial astrophysics scenario assumed in the
left figure.}
\label{fig:ConstraintsPrediction}
\end{figure*}

Estimating the impact of the constraints on the unresolved CXB on the X-ray spectrum in Eq.~\eqref{eq: averaged photon spectrum X} involves the modelling of the X-ray spectrum above $E_{\text{max}}$. While Ref.~\cite{Fialkov:2016zyq} assumed a single power law, extrapolating Eq.~\eqref{eq: averaged photon spectrum X} up to $14\text{ keV}$, here we assume a second power law above $3.5\text{ keV}$ taking the softest choice consistent with simulations of HMXBs (corresponding to $\beta_\text{X}\approx -3.2$), and matching it to the soft X-ray spectrum in Eq.~\eqref{eq: averaged photon spectrum X}. This procedure gives a conservative constraint on the X-ray flux assuming HMXBs are the dominant X-ray sources during the Cosmic Dawn.


\subsection{The neutral hydrogen fraction}\label{sec:ion}
During Cosmic Dawn, astrophysical objects emit X-ray and UV photons, reionizing the universe by $z\sim6$~\cite{McGreer:2014qwa}. In our 21-cm code we model the X-ray and UV ionization following Refs.~\cite{madau1999radiative,Mason_2019}~\footnote{A full inhomogeneous treatment would require the use of intricate simulations (see~\cite{so2014fully,Gnedin_2014,Iliev_2014,Trac_2015,Finlator_2018,Kannan_2021} for examples of full numerical simulations, and \cite{Mesinger:2010ne} for the semi-analytical simulation applied in the public 21cmFAST code). However, despite its simplicity, upon comparing our results with those obtained from the semi-analytical code 21cmFAST~\cite{Mesinger:2010ne}, we observe a high level of accuracy in reproducing the global 21-cm signal for redshifts $z\gtrsim10$ as well as the electron scattering optical depth to CMB.}, correcting for the total number of UV ionizing photons according to~\cite{Park:2018ljd}. The latter assumes that the fraction of ionizing photons which escape the ionized regions surrounding the emitting source is a power law in halo mass, as suggested by high redshift simulations~\cite{Paardekooper_2015, Xu_2016}.

Planck measurements of large scale CMB anisotropies measure the optical depth to reionization, $\tau_e$, yielding  $\tau_{e}= 0.054\pm0.0070$ at $68\%$ C.L.~\cite{Planck:2018vyg}. This measurement can be mapped to a constraint on the ionized fraction $x_e$ (or equivilantly the neutral fraction of hydrogen $x_{HI}$) according to the relation 
\begin{equation}
\tau_{e}=n_{H}^0 \sigma_T\int^{50}_{0} d z x_e(z)\frac{(1+z)^2}{H(z)}\, ,\label{eq:opticaldepth}
\end{equation}
where $n_{HI}^0$ is the number density of hydrogen atoms today, $\sigma_T$ is the Thompson cross section, and the upper integration limit is chosen to be high enough to capture the full contribution to $\tau_{e}$. In addition, missing \Lya and Ly-$\beta$ photons in bright quasars spectra can also be used as a probe for reionization, setting a 1$\sigma$ upper limit of $x_{HI}<0.06+0.05$ at $z=5.9$~\cite{McGreer:2014qwa}.

Since reionization is generally led by UV photons, the Planck measurement provides an upper bound on the UV photons that escape their host galaxies.

\section{Results}~\label{sec:results}

We are now prepared to explore the possibilities of leveraging the 21-cm global signal 
as a 
tool to unravel insights about the 2cDM dark sector.
Specifically, we focus on the 
DM implications 
for the minimal temperature at Cosmic Dawn alone, leaving the utilization of the full spectral shape, which entails greater constraining and discovery powers,  for the upcoming publication~\cite{our}. 

Without delving into the complicated systematic uncertainties which affect the 21-cm measurements, we  entertain here as an illustration the possibility of a non-detection of the Cosmic Dawn peak with sensitivity of $-100\units{mK}$. 
This assumption enables us to gauge the anticipated sensitivity to new physics in the global 21-cm spectrum, taking into account  the astrophysical uncertainties.

To gear our expectations we first adopt the fiducial astrophysical model of Ref.~\cite{Park:2018ljd} as our benchmark model with $F_\star = 0.05$, $\alpha_\star = 0.5$, $M_{\rm cut}=5\times10^8M_\odot$ $f_{\rm X} = 1$, $E_{\rm min}=0.5$, $f_{\rm esc}=0.1$, $\alpha_{\rm esc}=-0.5$, $t_\star = 0.5$. Within this model we then evaluate the amplitude and redshift of the Cosmic Dawn absorption signal in the mDM mass, $m_\text{m}$, and charge, $Q$ plane. The results are shown on the left of  Fig.~\ref{fig:ConstraintsPrediction}, where we see that most of the allowed parameter space with mDM mass below 10 GeV predicts an absoprtion dip which is deeper than the minimal one in standard cosmology.   Furthermore, with sensitivity of $-100$ mK, most of the viable parameter space for the above astrophysics scenario, could be excluded.  

In order to asses the systematic uncertainties on these conclusions, we perform a wide scan on the astrophysical parameters (working within the power-law SFR framework discussed in~\cite{Bouwens_2015,Park:2018ljd})
together with the dark sector parameters $(m_\text{m},Q)$, systematically imposing the observational constraints discussed in Sec.~\ref{sec:astro}. 
We also evaluate the impact of the expected constraints on the unresolved CXB from Athena~\cite{Marchesi:2020smf}. The envelope of all possible  brightness temperatures after imposing the existing constraints is shown in Fig.~\ref{fig:T21_of_z_scan}.\footnote{In detail we take $\log_{10} F_\star\in[-2.09, -0.66]$, $\alpha_\star \in [0.29,0.70]$, $\log_{10} M_{\text{cut}}\in[7.91,9.43]$, $\log_{10}f_X \in [-2,2.1]$, $E_{\rm min} \in [0.19,0.85]$, $\alpha_{\rm esc} \in [-1.74,1.18]$, $\log_{10} f_{\rm esc} \in [-1.94,0.35]$, and keep $t_\star=0.5$ constant given its strong degeneracy with $F_\star$ (see Ref.~\cite{Park:2018ljd}).} 

The possible variations of the signal, associated with the astrophysical uncertainties, can now be mapped onto the 2cDM parameter space, as shown on the right of Fig.~\ref{fig:ConstraintsPrediction}.  While for any point in the $m_{\rm m}$-$Q$ plane there exists a viable astrophysics model for which the Cosmic Dawn absorption feature is below  $-100$ mK, in part of the parameter space, such a minimum occurs for {\it any} viable astrophysics model.  The dark purple region in the plot shows this part of the parameter space, scanning over all 
 astrophysical models which satisfy the constraints of Sec.~\ref{sec:astro}, and taking into account only the existing Chandra X-ray limits. Within this region, a null measurement with an experimental sensitivity of $-100\units{mK}$ in the 21-cm brightness temperature would result in a robust constraint on millicharged dark matter, \emph{independently} of astrophysics.  The lighter blue region shows the expected improvement once the limits from Athena will become available. For comparison, the yellow line corresponding to the the contour of $\min\left(T_{21}\right) = -100\units{mK}$ in the astrophysical benchmark of the left panel is also shown. Comparing the yellow line with the dark purple region shows the importance of varying the astrophysical parameters to account for degeneracies with dark sector physics.  
 Furthermore, we find that even under our very conservative assumption, with future X-ray measurements one could exclude this benchmark astrophysics model within standard cosmology, independently of any new physics scenario. 

We note that the expected sensitivity in Fig.~\ref{fig:ConstraintsPrediction} should be complemented with data from 21-cm interferometry which have the potential of further reducing the astrophysical degeneracies (see Ref.~\cite{Barkana:2022hko} for preliminary work in this direction). The study performed so far demonstrates that 21-cm measurements can  undoubtedly 
play a crucial role in closing the gap in the millicharge DM parameter space, in a complementary manner  to future accelerators searches~\cite{Harnik:2019zee,Magill:2018tbb,Berlin:2018bsc,Magill:2018tbb,Haas:2014dda,Ball:2016zrp,Kelly:2018brz,Harnik:2019zee}, low-threshold direct detection experiments above ground~\cite{Emken:2019tni} and ion traps~\cite{Budker:2021quh,Berlin:2023zpn}.

\section{Acknowledgments}
We thank Andrey Mesinger, Julian Munoz and Yuxiang Qin for illuminating discussions and feedback on their code. We also thank Hongwan Liu, Andrea Caputo, Laura Lopez-Honorez and Ely Kovets for discussions. OZK thanks the Alexander Zaks scholarship for its support. The work of DR is partially financed by the  PRIN 2022 – Next Generation EU. The work of TV is supported, in part, by the Israel Science Foundation (grant No. 1862/21), by the Binational Science Foundation (grant No. 2020220), by the NSF-BSF (grant No. 2021780) and by the European Research Council (ERC) under the EU Horizon 2020 Programme (ERC- CoG-2015, Proposal No. 682676 LDMThExp).

{\bf Declaration of generative AI and AI-assisted technologies in the writing process:}
During the preparation of this work the author(s) minimally used ChatGPT in order to improve the readability of a few paragraphs. After using this service, the author(s) reviewed and edited the content as needed and take(s) full responsibility for the content of the publication.

\bibliography{bib21.bib}

\end{document}